\documentclass[12pt]{article}
\usepackage{graphicx,psfrag,epsf}

\def\bm#1{\mbox{\boldmath$#1$\unboldmath}}

\newcommand{\MS}{$\overline{\mathrm{MS}}$\ }
\newcommand{\LMS}{\Lambda_{\overline{\mathrm{MS}}}}
\newcommand{\LQCD}{\Lambda_\mathrm{QCD}}
\newcommand{\bo}{\beta_0}

\newcommand\npb[3]{Nucl.\ Phys.\ B {\bf #1} (#2) #3}

\newcommand\plb[3]{Phys.\ Lett.\ B {\bf #1} (#2) #3}
\newcommand\prd[3]{Phys.\ Rev.\ D {\bf #1} (#2) #3}
\newcommand\prep[3]{Phys.\ Rep.\ {\bf #1} (#2) #3}
\newcommand\prl[3]{Phys.\ Rev.\ Lett.\ {\bf #1} (#2) #3}
\newcommand\rmp[3]{Rev.\ Mod.\ Phys.\ {\bf #1} (#2) #3}

\newcommand{\hepph}[1]{{\tt hep-ph/#1}}

\begin{document}

\begin{titlepage}

\begin{flushright}
CLNS~02/1777\\
{\tt hep-ph/0202128}\\[0.2cm]
February 2002
\end{flushright}

\vspace{1.2cm}
\begin{center}
\Large\bf\boldmath
Higher-Order Corrections to QCD Factorization in $B\to\pi K,\pi\pi$ 
Decays\unboldmath
\end{center}

\vspace{0.5cm}
\begin{center}
Matthias Neubert and Ben D. Pecjak\\[0.1cm]
{\sl Newman Laboratory of Nuclear Studies, Cornell University\\
Ithaca, NY 14853, USA}
\end{center}

\vspace{0.8cm}
\begin{abstract}
\vspace{0.2cm}\noindent 
The renormalon calculus is used to calculate the terms of order 
$\beta_0^{n-1}\alpha_s^n$ in the perturbative expansions of the Wilson 
coefficients and hard-scattering kernels entering the QCD factorization 
formula for hadronic $B$-meson decays into two light pseudoscalar 
mesons. The asymptotic behavior of the expansions is analyzed, and a 
minimal model of power corrections arising from soft ``non-factorizable''
gluon exchange to the $B\to\pi K,\pi\pi$ decay amplitudes is obtained, 
which takes into account the structure of the leading and subleading 
infrared renormalon singularities. Whereas the resulting power 
corrections are generally very small, some of the strong-interaction 
phases of the hard-scattering kernels receive sizeable two-loop 
corrections. The implications of these findings on CP asymmetries and
branching ratios are investigated.
\end{abstract}

\vspace{1.0cm}
\centerline{\sl (Published in Journal of High-Energy Physics, 
JHEP02 (2002) 028)}

\end{titlepage}

\section{Introduction}

A theoretical understanding of hadronic $B$ decays is necessary to 
extract CKM matrix elements and CP-violating weak phases from data taken 
at the $B$ factories. The formalism of QCD factorization developed in 
\cite{BBNS1,BBNS2,BBNS3} has shown that two-body hadronic $B$ decays 
simplify considerably in the heavy-quark limit. In $B\to\pi K$ decays, 
for example, QCD factorization states that the matrix elements of the 
local operators $Q_i$ appearing in the effective weak Hamiltonian obey 
the factorization formula
\begin{eqnarray}
   \langle\pi K|Q_i|B\rangle 
   &=& F_0^{B\to\pi}\,T_{K,i}^I\otimes f_K\Phi_K
    + F_0^{B\to K}\,T_{\pi,i}^I\otimes f_\pi\Phi_\pi \nonumber\\
   &&\mbox{}+ T_i^{II}\otimes f_B\Phi_B\otimes f_K\Phi_K\otimes 
    f_\pi\Phi_\pi \,, 
\end{eqnarray} 
where $\Phi_M$ are leading-twist light-cone distribution amplitudes, 
and the $\otimes$-products imply integration over the light-cone 
momentum fractions of the constituent quarks inside the mesons. The 
hard-scattering kernels $T_i$ are perturbatively calculable quantities. 
The factorization formula is true to all orders in perturbation theory,
but it receives power corrections in $\LQCD/m_b$. If QCD factorization 
is to be used as a tool for extracting the flavor parameters of the 
Standard Model, it is important to show that these power corrections 
are not anomalously large. Some sources of power corrections were 
identified and estimated in \cite{BBNS2,BBNS3,Khodjamirian:2001mi}. In 
\cite{Burrell:2001pf,BNP}, the renormalon calculus was used to study 
power corrections due to soft gluon exchange in the decays 
$B\to D^{(*)} L$, where $L$ is a light meson. Here we generalize and 
extend these analyses to  focus on soft gluon corrections to the 
amplitudes for $B$ decays into two light pseudoscalar mesons. 

The model of power corrections that we develop relies on the analysis 
of infrared (IR) renormalon singularities, which arise because the 
low-momentum regions in loop diagrams give increasingly important 
contributions at higher orders in perturbation theory 
\cite{Beneke:1994qe,Neubert:1994vb,beneke1}, rendering perturbation 
theory a divergent asymptotic expansion. IR renormalons are most 
transparently studied through singularities in the Borel transforms of 
perturbation series with respect to the coupling constant \cite{tHof}. 
The Borel transform $S(u)$ of the perturbation series for a physical 
quantity has poles (or branch cuts) at half-integer values of the 
Borel variable $u$, which lead to ambiguities of order $(\LQCD/Q)^{2u}$ 
when performing the inverse transform, a process to which we will refer 
as Borel resummation. (Here $Q$ is the large momentum scale in the 
process.) Since the final answer for a physical quantity must be 
unambiguous, these renormalon ambiguities should be combined with power 
corrections of the same order in $1/Q$. Therefore, an analysis of IR 
renormalons may reveal non-trivial information about power corrections 
to various processes, which is particularly useful in cases where such 
corrections cannot be analyzed using an operator product expansion (see 
\cite{MaWe95,Web94,Akh95,Nason,KoSt95,Mar2} for early applications of 
this idea).

We calculate the Borel transforms of the hard-scattering kernels 
appearing in the QCD factorization formula using the so-called 
``large-$\bo$ limit'', in which the momentum flow through gluon lines 
is traced by inserting fermion bubbles on the propagators, replacing 
$n_f\to-3\bo/2$ at the end of the calculation \cite{beneke1}. In 
this way, the diagrams are effectively computed using a 
momentum-dependent running coupling at the vertices. This procedure 
gives the dominant higher-order contributions to a perturbation 
series in the very formal limit where $n_f\to-\infty$, in which case 
$\bo\to\infty$ with fixed number of colors. Although for real QCD 
$\bo=11-\frac23 n_f$ is not a large parameter, the use of the 
large-$\bo$ limit is motivated by the empirical observation that
perturbation series are often dominated by  terms of the form 
$\bo^{n-1}\alpha_s^n$. 

In this paper, we calculate both the Wilson coefficients and the 
hard-scattering kernels entering the QCD factorization formula at 
next-to-leading order (NLO) in the large-$\bo$ limit. This is 
necessary for obtaining renormalization-group (RG) invariant expressions 
for the physical decay amplitudes. The expansion of the Wilson 
coefficients in the large-$\bo$ limit is a novel feature of our work, 
which goes beyond the approach taken in \cite{Burrell:2001pf,BNP}. The 
numerical study in Section~\ref{numerical} helps to elucidate the 
main results of our work. We find that the power corrections to QCD 
factorization due to soft-gluon exchange are small, typically of 
order $(\LQCD/m_b)^2$. The only exception is a first-order power 
correction to the hard-spectator term, which is to be expected 
considering that the effective scale for gluon exchange with the 
spectator quark is of order $\sqrt{\LQCD m_b}$ \cite{BBNS3}. Perhaps
our most important result is that the strong-interaction phases of the 
decay amplitudes increase significantly when going beyond one-loop 
order in perturbation theory. This can  potentially enhance the 
direct CP asymmetries in decays such as $B\to\pi\pi$ and $B\to\pi K$ 
with respect to the predictions found at NLO in RG-improved perturbation 
theory in \cite{BBNS3}. Whether or not such increased asymmetries are 
realistic depends on whether there are significant multi-loop 
contributions not captured in the large-$\bo$ limit. We leave a more 
careful analysis of such contributions for future work.

\boldmath
\section{Wilson coefficients in the large-$\bo$ limit} 
\unboldmath
\label{sec-Wilson}

The QCD factorization formula provides model-independent expressions 
for hadronic weak decay amplitudes at leading power in the heavy-quark 
expansion. All non-trivial strong-interaction effects, which give 
corrections to the so-called ``naive'' factorization (vacuum insertion)
approximation, are contained in  RG-invariant parameters $a_i$, which 
are  products of Wilson coefficients with convolutions of 
hard-scattering kernels and meson light-cone distribution amplitudes. 
In order to obtain RG-invariant results, it is crucial that the Wilson 
coefficients and hard-scattering kernels be treated in  the same 
perturbative approximation scheme. Since our goal is to resum the 
perturbation series for the hard-scattering kernels in the large-$\bo$
limit, we must first compute the Wilson coefficients appearing in the
effective weak Hamiltonian in the same limit. This is non-trivial, since
the counting scheme underlying the large-$\bo$ limit is different from 
the usual counting of logarithms used in the calculation of the Wilson
coefficients. Specifically, whereas $\alpha_s\ln(M/\mu)$ is considered 
$O(1)$ in the usual RG counting, this quantity is $O(1/\bo)$ in the 
large-$\bo$ limit. On the contrary, whereas $1/\ln(M/\mu)$ is usually 
treated as a small parameter, it is considered $O(1)$ in the large-$\bo$ 
limit. 

Consider, as an example, the rare hadronic decays $B\to\pi\pi$, for 
which the effective weak Hamiltonian (neglecting small contributions
from electroweak penguins) is \cite{Buchalla:1995vs}
\begin{equation}
   {\cal H}_{\rm eff} = \frac{G_F}{\sqrt2} \sum_{p=u,c}\,\lambda_p
   \bigg( \sum_{i=1,2}\,C_i(\mu)\,Q_i^p 
   + \sum_{i=3,\dots,6} C_i(\mu)\,Q_i + C_{8g}(\mu)\,Q_{8g} \bigg) ,
\end{equation}
where $\lambda_p=V_{pb} V_{pd}^*$ are products of CKM matrix elements.
The scale dependence of the Wilson coefficients is governed by the 
evolution equation (in compact matrix notation)
\begin{equation}\label{evolution}
   \vec C(\mu) = T_\alpha\,\exp\Bigg[
   - \int\limits_{\alpha_s(M)}^{\alpha_s(\mu)}\!d\alpha\,
   \frac{\bm{\gamma}^T(\alpha)}{2\alpha\,\beta(\alpha)} \Bigg]\,
   \vec C(M) \,,
\end{equation}
where the evolution matrix is given in terms of an ordered matrix 
exponential involving the anomalous dimension matrix $\bm{\gamma}$ and 
the $\beta$-function. The initial values $\vec C(M)$ at a scale 
$M\sim m_W$ are obtained by matching the Standard Model onto a 
low-energy effective theory, in which the heavy fields of the 
electroweak gauge bosons and the top quark are integrated out. 

To evaluate (\ref{evolution}) at subleading order in the large-$\bo$ 
limit, we expand the QCD $\beta$-function as $\beta(\alpha)=b+O(1/\bo)$, 
where the rescaled coupling
\begin{equation}
   b(\mu) = \bo\,\frac{\alpha_s(\mu)}{4\pi}
   = \frac{1}{\ln(\mu^2/\LMS^2)}
\end{equation}
is $O(1)$ in the large-$\bo$ limit. Next, we change variables from 
$\alpha$ to $b$ in the integral in (\ref{evolution}), write the 
anomalous dimension matrix as a function of $b$, and expand in powers
of $1/\bo$. Taking into account that factors of $n_f$ must be replaced 
with $-3\bo/2$, we find that in $2\times 5$ block form the resulting 
matrix is given by
\begin{equation}
   \bm{\gamma}^T(b) = \left( \begin{array}{c|c}
    0 & 0 \\\hline 
    ~0~ & ~b\,\bm{C_0} \end{array} \right)
   + \frac{1}{\bo}
   \left( \begin{array}{c|c}
    \bm{A_1}(b)~ & 0 \\\hline
    b\,\bm{B_1} & ~\bm{C_1}(b) \end{array} \right) + O(1/\bo^2) \,.
\end{equation}
The justification for writing the leading term in this expansion as a 
constant $5\times 5$ matrix $\bm{C_0}$ times the coupling constant $b$ 
is that this corner of the anomalous dimension matrix does not pick up 
an additional factor of $n_f$ when going from leading order (LO) to 
next-to-leading order in perturbation theory, so that all further 
perturbative contributions are absorbed in the matrix $\bm{C_1}(b)$, 
the form of which is irrelevant to our discussion. A similar argument 
applies to the submatrix in the lower left corner, which at order 
$1/\bo$ is given by a constant $5\times 2$ matrix $\bm{B_1}$ times $b$. 

The matrix $\bm{A_1}(b)$ governing the evolution of the 
current--current operators $Q_{1,2}^p$ in the large-$\bo$ limit is a
non-trivial function of the coupling, which has been calculated in 
\cite{BNP,pott}. The result is
\begin{equation}
   \bm{A_1}(b) = \gamma(b) \left( \begin{array}{cc}
    -\frac{1}{N}~ & 1\\
    1 & -\frac{1}{N} 
   \end{array}\right) ,
\end{equation}
where $N=3$ is the number of colors, and
\begin{equation}
   \gamma(b) = \frac{2b(3+2b)(1-b)\,\Gamma(4+2b)}
        {3\Gamma(1-b)\,\Gamma^2(2+b)\,\Gamma(3+b)} \,.
\end{equation}
The matrices $\bm{C_0}$ and $\bm{B_1}$ can be derived from the 
expressions for the LO anomalous dimension matrices given in 
\cite{Buchalla:1995vs}:
\begin{equation}
{\arraycolsep=0.15cm
   \bm{C_0} = \left( \begin{array}{ccccc}
    0 & \frac{1}{N} & 0 & \frac{1}{N} & 0 \\
    0 & -1 & 0 & -1 & 0 \\
    0 & \frac{1}{N} & 0 & \frac{1}{N} & 0 \\
    0 & -1 & 0 & -1 & 0 \\
    -\frac92 & -\frac{11N}{6}\!+\!\frac{29}{6N} & \frac92 &
     \frac{8N}{3}\!-\!\frac{25}{6N} & 0
   \end{array} \right) , 
   \qquad
   \bm{B_1} = \left( \begin{array}{cc}
    -\frac{2}{3N} & 0 \\
    \frac23 & 0 \\ 
    -\frac{2}{3N} & 0 \\
    \frac23 & 0 \\
    \frac{11N}{9}\!-\!\frac{29}{9N} & 3 
   \end{array}\right) .}
\end{equation}
Taking into account the textures of these matrices as well as 
corresponding relations between the matching conditions for the various 
coefficients at the scale $\mu=m_W$, we find that the solutions for the 
Wilson coefficients can be expanded as
\begin{eqnarray}
   C_1(\mu) &=& 1 + \frac{1}{\bo}\,C_1^{(1)}(\mu) + O(1/\bo^2) \,, 
    \nonumber\\
   C_i(\mu) &=& \frac{1}{\bo}\,C_i^{(1)}(\mu) + O(1/\bo^2) \,;
    \quad i=2,\dots,6 \,, \nonumber\\
   C_{8g}(\mu) &=& C_{8g}^{(0)}(\mu) + \frac{1}{\bo}\,C_{8g}^{(1)}(\mu)
    + O(1/\bo^2) \,,\end{eqnarray}
where 
\begin{equation}
   C_2^{(1)}(\mu) = - N\,C_1^{(1)}(\mu) \,, 
   \qquad
   C_4^{(1)}(\mu) = C_6^{(1)}(\mu) = - N\,C_3^{(1)}(\mu)
    = - N\,C_5^{(1)}(\mu) \,.
\end{equation}
Up to $O(1/\bo)$, the Wilson coefficients $C_1,\dots,C_6$ are thus 
determined in terms of two functions, given by
\begin{eqnarray}\label{C4evol}
   C_2^{(1)}(\mu) &=& C_2^{(1)}(M) - \int\limits_{b(M)}^{b(\mu)}\!db\,
    \frac{\gamma(b)}{2b^2} \,, \nonumber\\
   C_4^{(1)}(\mu) &=& C_4^{(1)}(M)
    + \bigg( \frac{b(\mu)}{b(M)} - 1 \bigg)
    \left( C_4^{(1)}(M) - \frac13 \right) .
\end{eqnarray}
The coefficient of the chromo-magnetic operator $Q_{8g}$ will be needed 
only at LO, where it is scale independent, i.e.,
\begin{equation}
   C_{8g}^{(0)}(\mu) = C_{8g}^{(0)}(M) \,.
\end{equation}

Note that, as a consequence of the fact that the matrix $\bm{C_0}$ has
eigenvalues 0 and $-2$, the scale dependence of the penguin coefficients
$C_3,\dots,C_6$ governed by the function $C_4^{(1)}(\mu)$ is extremely
simple, and given by $b(\mu)/b(M)-1=2b(\mu)\ln(M/\mu)$. We will see 
below that the appearance of a logarithm is indeed required to cancel 
the scale dependence of the hard-scattering kernels associated with 
penguin contractions of the operators $Q_i$.

\begin{table}
\centerline{\parbox{15cm}{\caption{\label{tab-numwil}
Comparison of the Wilson coefficients in the large-$\bo$ limit with 
those obtained at NLO in conventional, RG-improved perturbation theory. 
Input parameters are $m_t(m_t)=167$\,GeV, $m_b(m_b)=4.2$\,GeV, 
$\alpha_s(m_b)=0.2244$, and $\alpha_s(m_W)=0.1202$.}}}
\vspace{0.2cm}
\begin{center}
\begin{tabular}{||l|c|c|c|c|c|c|c||} 
\hline\hline
NLO & $C_1$ & $C_2$ & $C_3$ & $C_4$ & $C_5$ & $C_6$ \\ 
\hline
$\mu=m_b/2$ & 1.137 & $-0.295$ & 0.021 & $-0.051$ & 0.010 & $-0.065$ \\ 
$\mu=m_b$   & 1.081 & $-0.190$ & 0.014 & $-0.036$ & 0.009 & $-0.042$ \\ 
$\mu=2m_b$  & 1.045 & $-0.113$ & 0.009 & $-0.025$ & 0.007 & $-0.027$ \\ 
\hline\hline
Large-$\bo$ limit & $C_1$ & $C_2$ & $C_3$ & $C_4$ & $C_5$ & $C_6$ \\ 
\hline
$\mu=m_b/2$ & 1.087 & $-0.260$ & 0.020 & $-0.061$ & 0.020 & $-0.061$ \\
$\mu=m_b$   & 1.061 & $-0.183$ & 0.014 & $-0.041$ & 0.014 & $-0.041$ \\
$\mu=2m_b$  & 1.039 & $-0.118$ & 0.009 & $-0.028$ & 0.009 & $-0.028$ \\ 
\hline\hline
\end{tabular}
\end{center}
\end{table}

The remaining  step in the calculation of the Wilson coefficients is 
to specify the matching conditions at a scale $M\sim m_W$. In general, 
these will be complicated functions of the coupling $b(m_W)$, which is 
formally $O(1)$ in the large-$\bo$ limit. However, since 
$b(m_W)\approx 0.07$ is numerically small, it will be a good 
approximation to keep only the one-loop contributions to the matching 
conditions. They read (in the NDR scheme)
\begin{equation}
   C_2^{(1)}(m_W) \approx \frac{11}{2}\,b(m_W) \,, 
   \quad
   C_4^{(1)}(m_W) \approx \frac12\,\widetilde E_0(x_t)\,b(m_W) \,,
   \quad
   C_{8g}^{(0)}(m_W) \approx F(x_t) \,,
\end{equation}
where $x_t=(m_t/m_W)^2$, and
\begin{eqnarray}
   \widetilde E_0(x) &=& - \frac23(1+\ln x)
    + \frac{x(18-11x-x^2)}{12(1-x)^3}
    + \frac{x^2(15-16x+4x^2)}{6(1-x)^4}\ln x \,, \nonumber\\
   F(x) &=& - \frac{x(2+5x-x^2)}{8(1-x)^3}
    - \frac{3x^2}{4(1-x)^4} \ln x \,.
\end{eqnarray}
In Table~\ref{tab-numwil}, we compare our results for the Wilson 
coefficients obtained at subleading order in the large-$\bo$ limit with
those derived using the standard RG evolution at NLO. Given the great 
simplifications that occur in the large-$\bo$ limit, it is remarkable
that the numerical values obtained in the two schemes are very close
to each other.

\section{Borel resummation for the decay amplitudes}
\label{sec-borel}

\begin{figure}
\centerline{\epsfxsize=10cm\epsffile{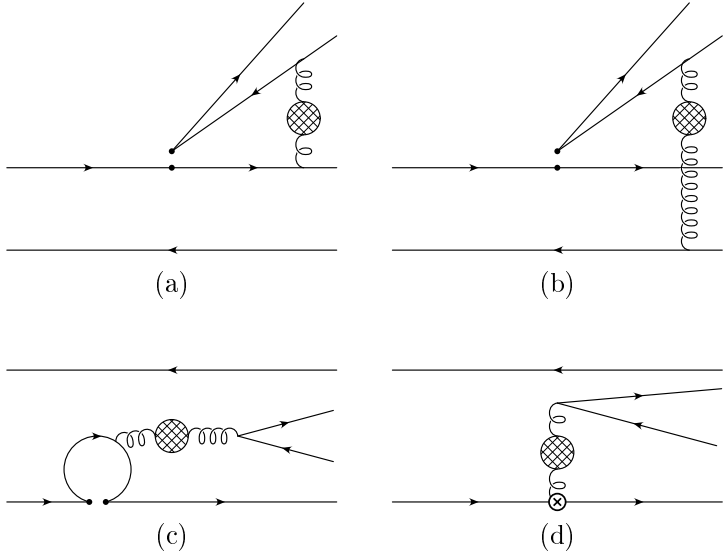}}
\centerline{\parbox{15cm}{\caption{\label{fig:graphs}
Representative examples of contributions to the hard-scattering kernels 
in the QCD factorization formula including bubble resummation of the 
gluon propagators: (a) vertex contribution, (b) hard-scattering with the
spectator quark, (c) penguin contribution, (d) chromo-magnetic dipole
contribution.}}}
\end{figure}

Having obtained expressions for the Wilson coefficients in the 
large-$\bo$ limit, our next task is to calculate the hard-scattering
kernels in the QCD factorization formula using the same resummation 
scheme. There are three different types of kernels, corresponding to 
``non-factorizable'' vertex corrections, ``non-factorizable'' gluon 
exchange with the spectator quark in the $B$ meson, and penguin 
contributions. Representative examples of each kind are depicted in 
Figure~\ref{fig:graphs}, where (c) and (d) are both part of the penguin
contribution. The weak decay amplitudes are obtained by evaluating the 
hadronic matrix elements of a ``factorized transition operator'' 
consisting of products of the parameters $a_i$ and two quark currents, 
whose matrix elements are known in terms of meson decay constants and 
semi-leptonic form factors \cite{BBNS2,BBNS3}. As an example, we quote 
the resulting expressions for the $B\to\pi\pi$ decay amplitudes. They 
are
\begin{eqnarray}\label{Bpipi}
   -{\cal A}(\bar B^0\to\pi^+\pi^-)
   &=& \bigg[ \lambda_u\,a_1 + \sum_{p=u,c} \lambda_p\,
    (a_4^p + r_\chi\,a_6^p) \bigg] A_{\pi\pi} \,,
    \nonumber\\[-0.2cm]
   -\sqrt2\,{\cal A}(B^-\to\pi^-\pi^0)
   &=& \lambda_u\,(a_1 + a_2)\,A_{\pi\pi} \,, \nonumber\\[0.2cm]
   {\cal A}(\bar{B}^0\to\pi^0\pi^0)
   &=& \sqrt2\,{\cal A}(B^-\to\pi^-\pi^0)
    - {\cal A}(\bar B^0\to\pi^+\pi^-) \,,
\end{eqnarray}
where
\begin{equation}
   A_{\pi\pi} = i\frac{G_F}{\sqrt2}\,(m_B^2-m_\pi^2)\,
    F_0^{B\to\pi}(m_\pi^2)\,f_\pi \,,
   \qquad
   r_\chi = \frac{2m_\pi^2}{m_b(m_u+m_d)} \,, 
\end{equation}
and we have neglected small electroweak penguin contributions and 
power-suppressed weak annihilation terms. Similar results hold for 
other decays such as $B\to\pi K$. The general expressions for the $a_i$ 
parameters at leading order in the heavy-quark expansion read (we quote 
results for $B\to\pi K$ decays; those for $B\to\pi\pi$ are obtained by 
replacing $K\to\pi$ everywhere)
\begin{eqnarray}\label{aiBBNS}
   a_1 &=& C_1(\mu) + \frac{C_2(\mu)}{N} \left[ 1
    + \frac{C_F\alpha_s(\mu)}{4\pi}\,V_K(\mu)
    + \frac{C_F\pi\alpha_s(\mu)}{N}\,H_{K\pi}(\mu) \right] ,
    \nonumber\\
   a_2 &=& C_2(\mu) + \frac{C_1(\mu)}{N} \left[ 1
    + \frac{C_F\alpha_s(\mu)}{4\pi}\,V_\pi(\mu)
    + \frac{C_F\pi\alpha_s(\mu)}{N}\,H_{\pi K}(\mu) \right] , 
    \nonumber\\
   a_4^p &=& C_4(\mu) + \frac{C_3(\mu)}{N} \left[ 1
    + \frac{C_F\alpha_s(\mu)}{4\pi}\,V_K(\mu)
    + \frac{C_F\pi\alpha_s(\mu)}{N}\,H_{K\pi}(\mu) \right]
    + \frac{C_F\alpha_s(\mu)}{4\pi}\,\frac{P_{K,2}^p(\mu)}{N} \,,
    \nonumber\\
   a_6^p &=& C_6(\mu) + \frac{C_5(\mu)}{N} \left[ 1
    -6\cdot\frac{C_F\alpha_s(\mu)}{4\pi} \right] 
    + \frac{C_F\alpha_s(\mu)}{4\pi}\,\frac{P_{K,3}^p(\mu)}{N} \,.
\end{eqnarray}
The quantities $V$, $H$ and $P$ include the NLO corrections from the 
vertex, hard-spectator and penguin diagrams, respectively \cite{BBNS3}. 
The subscript $n=2,3$ on $P_{K,n}^p$ refers to the twist of the 
corresponding light-cone distribution amplitude of the kaon. 

\subsection{General results for the Borel transforms} 

The summation of terms of order $\bo^{n-1}\alpha_s^n$ in the kernels 
$V$, $H$ and $P$ is achieved by repeating the calculations of 
\cite{BBNS3} inserting an arbitrary number of fermion loops on the gluon 
propagators (see \cite{BNP} for details). The results are then written 
as integrals over the Borel variable $u$. Using the fact that 
$\alpha_s=O(1/\bo)$ and the explicit results for the Wilson coefficients 
obtained in the previous section, it follows that
\begin{equation}\label{a1fin}
   a_1 = 1 + O(1/\bo^2)
\end{equation}
becomes trivial in the large-$\bo$ limit. Also, the parameters $a_3$ 
and $a_5$, which do not enter the $B\to\pi\pi$ decay amplitudes in 
(\ref{Bpipi}), vanish to this order, i.e., $a_3,\,a_5=O(1/\bo^2)$. 
These results compare well with the full NLO analysis in \cite{BBNS3}, 
where it was found that $a_1\approx 1.03+0.02i$, while the values of 
$a_3$ and $a_5$ are below 1\% in magnitude. 

The results for the three remaining coefficients are non-trivial. In the 
case of $a_2$, both the vertex term $V_\pi$ and the hard-spectator term 
$H_{\pi K}$ contribute. The vertex term has been computed in the 
large-$\bo$ limit in \cite{BNP}, with the result that
\begin{equation}
   \frac{\alpha_s(\mu)}{4\pi}\,V_\pi(\mu) 
   \to \frac{2}{\bo}\,\Bigg[ \int\limits_0^{b(\mu)}
    \frac{db}{b} \left( \frac{\gamma(b)}{2b} - 3 \right) + 
    3\ln\frac{b(\mu)}{b(m_b)} + \int\limits_0^\infty\!du\,e^{-u/b(m_b)}\,
   e^{5u/3}\,S_V(u) \Bigg] \,,
\end{equation}
and the Borel transform
\begin{eqnarray}\label{sV}
   S_V(u) &=& \int\limits_0^1\!dx\,\Phi_\pi(x) \Bigg[
    \frac{3}{u}\,e^{-5u/3} - \frac{(3-u)\,x^{-u}}{(1-u)(2-u)}
    (\pi\cot\pi u+i\pi) \nonumber\\
   &&\mbox{}- \frac{(3-u)\Gamma(2-2u)\Gamma(u)}{\Gamma(3-u)}
    \int\limits_0^1\!dz\,\frac{1}{\big[z(x+(1-x)z)\big]^u} \Bigg] .
\end{eqnarray}
Here and below we take advantage of the symmetry of the pion 
distribution amplitude under $x\leftrightarrow(1-x)$ to simplify the
results. The term proportional to $e^{-5u/3}$ subtracts the pole at 
$u=0$, as appropriate in the \MS scheme. In the large-$\bo$ limit, the 
hard-spectator contribution is obtained from bubble resummation on the 
gluon propagator of a tree diagram, so the result is scale independent 
and given by
\begin{equation}
   \frac{\pi\alpha_s(\mu)}{N}\,H_{\pi K}(\mu) 
   \to \frac{2}{\bo} \int\limits_0^\infty\!du\,e^{-u/b(m_b)}\,
   e^{5u/3}\,S_H(u) \,,
\end{equation}
where (we use the short-hand notation $\bar y\equiv 1-y$)
\begin{equation}
   S_H(u) = \frac{2\pi^2}{N}\,\frac{f_B f_K}{m_B^2\,F_0^{B\to K}(0)}
    \int\limits_0^1\!dx \int\limits_0^1\!dy \int\limits_0^1\!d\xi\,
    \frac{\Phi_\pi(x)\,\Phi_K(y)\,\Phi_B(\xi)}
         {x\,\big(\bar y\,\xi\big)^{1+u}} \,.
\end{equation}
Combining the various pieces, it is readily seen that the scale 
dependence cancels between the Wilson coefficients and the kernels. The
RG-invariant result is
\begin{eqnarray}\label{a2fin}
   a_2 &=& \frac{1}{N} + \frac{2 C_F}{N\bo}\,\Bigg\{
    C_2^{(1)}(m_b) + \int\limits_0^{b(m_b)} \frac{db}{b}
    \left( \frac{\gamma(b)}{2b} - 3 \right) \nonumber\\
   &&\qquad\mbox{}+ \int\limits_0^\infty\!du\,e^{-u/b(m_b)}\,e^{5u/3}\,
    \Big[ S_V(u) + S_H(u) \Big] \Bigg\} + O(1/\bo^2) \,.
\end{eqnarray}

Let us now discuss the resummation for the penguin coefficients 
$a_{4,6}^p$. It follows from our earlier results that in this case the
vertex and hard-spectator contributions enter only at $O(1/\bo^2)$, and
so it suffices to calculate the penguin kernels $P_{K,n}^p$ in 
(\ref{aiBBNS}). The cancellation of the scale dependence between the 
Wilson coefficients and these kernels is subtle and relies in an 
essential way on our treatment of the Wilson coefficients in the 
large-$\bo$ limit. Consider the coefficient $a_4^p$ as an example. For 
fixed momentum fraction $y$, we find\footnote{For the sake of this 
perturbative argument it is legitimate to interchange the order of the 
integrations over $u$ and $y$.}
\begin{eqnarray}\label{a4p}
   \bo\,a_4^p(y) &=& C_4^{(1)}(\mu) + \frac{C_3^{(1)}(\mu)}{N}
    + \frac{C_F}{N} \int\limits_0^\infty\!du\,e^{-u/b(m_b)}\,
    e^{5u/3}\,\big(-\bar y-i\epsilon\big)^{-u} \nonumber\\
   &&\mbox{}\times \Bigg\{ \frac{n_f}{\bo}\,
    \Big[ C_4^{(1)}(\mu) + C_6^{(1)}(\mu) \Big]
    \left( \frac43\ln\frac{m_b}{\mu} - G(0,\bar y) \right) \nonumber\\
   &&\qquad\mbox{}+ C_1^{(0)}\,\bigg[ \frac43\ln\frac{m_b}{\mu}
    + \frac23 - G(s_p,\bar y) \bigg] - \frac{2C_{8g}^{(0)}}{\bar y}
    \Bigg\} + O(1/\bo) \,,
\end{eqnarray}
where $s_p=(m_p/m_b)^2$ for $p=u,c$, and
\begin{equation}\label{Gk}
   G(s,\bar y) = -4\int\limits_0^1\!dz\,z(1-z)
   \ln[s-z(1-z)\bar y-i\epsilon] \,.
\end{equation}
Noting that $G(0,\bar y)=\frac23[\frac53-\ln(-\bar y-i\epsilon)]$, 
replacing $n_f/\bo\to-3/2$, performing the integral over $u$, and using 
the results from Section~\ref{sec-Wilson}, we obtain
\begin{eqnarray}
   \bo\,a_4^p(y) &=& \frac{2C_F}{N} C_4^{(1)}(\mu) 
    + \frac{2C_F}{N}\,b(-\bar y\,e^{-5/3}\,m_b^2-i\epsilon) \nonumber\\
   &&\mbox{}\times \Bigg\{ C_4^{(1)}(\mu)
    \left( -2\ln\frac{m_b}{\mu} + \frac53 - \ln(-\bar y-i\epsilon) 
    \right) \nonumber\\
   &&\qquad\mbox{}+ \frac23\ln\frac{m_b}{\mu}
    + \frac13 - \frac12\,G(s_p,\bar y) - \frac{C_{8g}^{(0)}}{\bar y}
    \Bigg\} + O(1/\bo) \\
   &=& \frac{2C_F}{N}\,\Bigg\{ C_4^{(1)}(M) 
    + b(M) \bigg[ \frac13 + \frac12\,G(0,\bar y)
    - \frac12\,G(s_p,\bar y) - \frac{C_{8g}^{(0)}}{\bar y} \bigg]
    \Bigg\} + O(1/\bo) \,, \nonumber
\end{eqnarray}
where $M^2=-\bar y\,e^{-5/3}\,m_b^2-i\epsilon$, and in the last step we
have used the evolution equation for $C_4^{(1)}$ in (\ref{C4evol}). It 
follows that $a_4^p(y)$ is indeed independent of the scale $\mu$. The 
terms proportional to $C_4^{(1)}(\mu)+C_6^{(1)}(\mu)$ in (\ref{a4p}) 
are absorbed by setting the scale in the Wilson coefficients to be the 
(time-like) momentum $M$ carried by the gluon propagator. Having 
established that the resummed expression for $a_4^p$ is scale 
independent, we return to the original expression in (\ref{a4p}), set 
$\mu=m_b$, and integrate over the light-cone variable $y$ to obtain the 
final results
\begin{eqnarray}\label{a4a6fin}
   a_4^p &=& \frac{2C_F}{N\bo} \Bigg[ C_4^{(1)}(m_b)
    + \int\limits_0^\infty\!du\,e^{-u/b(m_b)}\,e^{5u/3}\,
    S_{P,2}^{p}(u) \Bigg]  + O(1/\bo^2) \,, \nonumber\\
   a_6^p &=& \frac{2C_F}{N\bo} \Bigg[ C_4^{(1)}(m_b)
    + \int\limits_0^\infty\!du\,e^{-u/b(m_b)}\,e^{5u/3}\,
    S_{P,3}^{p}(u) \Bigg]  + O(1/\bo^2) \,,
\end{eqnarray}
where
\begin{eqnarray}\label{SK23}
   S_{P,2}^{p}(u) &=& \int\limits_0^1\!dy\,\Phi_K(y)\,
    \frac{e^{i\pi u}}{\bar y^u} \Bigg[ \frac13 +
    \frac32\,C_4^{(1)}(m_b)\,G(0,\bar y) - \frac12\,G(s_p,\bar y)
    - \frac{C_{8g}^{(0)}}{\bar y} \Bigg] , \nonumber\\
   S_{P,3}^{p}(u) &=& \int\limits_0^1\!dy\,\Phi_K^p(y)\,
    \frac{e^{i\pi u}}{\bar y^u} \Bigg[ \frac13 +
    \frac32\,C_4^{(1)}(m_b)\,G(0,\bar y) - \frac12\,G(s_p,\bar y)
    - C_{8g}^{(0)} \Bigg] .
\end{eqnarray}

Equations (\ref{a1fin}), (\ref{a2fin}), and (\ref{a4a6fin}) are the main 
results of this work. They allow us to study the effects of higher-order 
$\bo^{n-1}\alpha_s^n$ corrections on the weak decay amplitudes for 
processes such as $B\to\pi\pi$ and $B\to\pi K$. Compared with the 
general expressions in (\ref{aiBBNS}), the terms that survive at NLO in 
the large-$\bo$ limit are rather simple. However, these terms still 
capture the bulk of the non-trivial strong-interaction effects in 
non-leptonic $B$ decays. In particular, it has already been noted in 
\cite{BBNS2,BBNS3} that $a_1$ is always very close to 1, that 
hard-spectator interactions are the  most important for the coefficient 
$a_2$, and that the penguin matrix elements give the dominant NLO
contributions to $a_{4,6}^p$. Before we evaluate our results 
numerically, we now simplify the expressions for the Borel transforms 
by using a model for the meson distribution amplitudes.

\subsection{Results with asymptotic distribution amplitudes} 
\label{subsec:asym}

The exact integral expressions for the Borel transforms of the 
hard-scattering kernels simplify if the asymptotic forms of the pion 
and kaon light-cone distribution amplitudes are adopted. They are
\begin{equation}
   \Phi_\pi(x) = 6x(1-x) \,, \qquad
   \Phi_K(y) = 6y(1-y) \,, \qquad
   \Phi_K^p(y) = 1 \,.
\end{equation}
In addition, for the hard-spectator term we employ a  simple model for 
the leading-twist distribution amplitude of the $B$-meson suggested in 
\cite{Grozin}, according to which 
\begin{equation}
   \Phi_B(\xi) = \frac{\xi}{\xi_0^2}\,e^{-\xi/\xi_0} \,;
   \quad \xi_0 = \frac{\lambda_B}{m_B} \,.
\end{equation}
The hadronic parameter $\lambda_B=O(\LQCD)$ is defined in terms of the 
first inverse moment, $\int_0^1(d\xi/\xi)\,\Phi_B(\xi)=m_B/\lambda_B$. 
Its value at present is rather uncertain, a typical range being 
$\lambda=(350\pm 150)$\,MeV.

The results obtained for the Borel transforms of the vertex and 
hard-spectator terms after integration over these distribution 
amplitudes are
\begin{eqnarray}\label{Spi}
   S_V(u) &=& \frac{3}{u}\,e^{-5u/3}
    - \frac{6\pi(\cot\pi u+i)}{(1-u)(2-u)^2} \nonumber\\
   &&\mbox{}- \frac{6\Gamma(1-2u)\,\Gamma(u)}{(2-u)\,\Gamma(3-u)}
    \left[ 1 + 2u(1-2u) \left( \psi(1-2u) - \psi(1-u) - \frac34 \right) 
    \right] , \nonumber\\
   S_H(u) &=& \frac{18\pi^2}{N}\,
    \frac{f_B f_K}{m_B\lambda_B\,F_0^{B\to K}(0)}\,
    \frac{2\Gamma(1-u)}{(1-u)(2-u)}
    \left( \frac{m_B}{\lambda_B} \right)^u ,
\end{eqnarray}
where $\psi(z)$ is the digamma function, defined as the logarithmic 
derivative of the gamma function. The results for the penguin 
contributions read
\begin{eqnarray}\label{SK23asy}
   S_{P,2}^{p}(u) &=& e^{i\pi u}\,\Bigg[
    \frac{2}{(2-u)(3-u)} + \frac32\,C_4^{(1)}(m_b)\,{\cal G}_K(0,u)
    - \frac12\,{\cal G}_K(s_p,u)
    - \frac{6C_{8g}^{(0)}}{(1-u)(2-u)} \Bigg] , \nonumber\\
   S_{P,3}^{p}(u) &=& e^{i\pi u}\,\Bigg[ \frac{1}{3(1-u)}
    + \frac32\,C_4^{(1)}(m_b)\,\widehat{\cal G}_K(0,u)
    - \frac12\,\widehat{\cal G}_K(s_p,u) - \frac{C_{8g}^{(0)}}{1-u}
    \Bigg] ,
\end{eqnarray}
where we have defined
\begin{equation}\label{penguins}
   {\cal G}_K(s,u) = \int\limits_0^1\!dy\,\Phi_K(y)\,\bar y^{-u}\,
    G(s,\bar y) \,, \qquad
   \widehat{\cal G}_K(s,u) = \int\limits_0^1\!dy\,\Phi_K^p(y)\,
   \bar y^{-u}\,G(s,\bar y) \,.
\end{equation}
For $p=u$, these functions can be evaluated in closed form, yielding
\begin{eqnarray}
   {\cal G}_K(0,u) &=& \frac{4}{(2-u)(3-u)} \left(
    \frac53 + i\pi + \frac{1}{2-u} + \frac{1}{3-u} \right) , \nonumber\\
   \widehat{\cal G}_K(0,u) &=& \frac{2}{3(1-u)} \left(
    \frac53 + i\pi + \frac{1}{1-u} \right) .
\end{eqnarray}
For the case of the charm-quark loop, the functions ${\cal G}_K(s_c,u)$ 
and $\widehat{\cal G}_K(s_c,u)$ with $s_c\ne 0$ do not have a simple 
analytic form. We choose instead to write them as infinite sums of poles 
in the Borel plane, by expanding the logarithm in (\ref{Gk}) in powers 
of $\bar y$ and then performing the integrals. We find that 
\begin{eqnarray}
   {\cal G}_K(s_c,u) &=& \frac{-4\ln s_c}{(2-u)(3-u)}
    + \sum_{n=1}^\infty\,\frac{\sqrt\pi}{2n}\,(4 s_c)^{-n}\,
    \frac{\Gamma(n+2)}{\Gamma(n+\frac52)}\,\frac{6}{(2+n-u)(3+n-u)} \,,
    \nonumber\\
   \widehat{\cal G}_K(s_c,u) &=& \frac{-2\ln s_c}{3(1-u)}
    + \sum_{n=1}^\infty\,\frac{\sqrt\pi}{2n}\,(4 s_c)^{-n}\,
    \frac{\Gamma(n+2)}{\Gamma(n+\frac52)}\,\frac{1}{1+n-u} \,.
\end{eqnarray}
The sums converge as long as $s_c>1/4$ and can be expressed in terms of 
hypergeometric functions. These functions can then be analytically 
continued to the physical region $s_c<1/4$, where they develop a 
non-zero imaginary part (recall that $s_c\to s_c-i\epsilon$). 

\subsection{IR renormalon ambiguities and power corrections}

Expansion of the Borel transforms (times $e^{5u/3}$) in powers of $u$ 
generates the perturbative series of the kernels in the \MS scheme. The
Borel sums of these series are obtained by performing the $u$-integrals 
exactly. However, the Borel transforms contain poles along the 
integration contour, which render the results of Borel summation 
ambiguous. This is the well-known problem of IR renormalons, which 
reflects the fact that the original perturbation series were asymptotic.

For a given Borel transform, the residue of the nearest pole provides a 
measure of the renormalon ambiguity, which is conventionally taken as 
an estimate of the power corrections that should be added to the 
perturbation series in order to get a well-defined result. We will now 
work out the structure of these leading singularities for the parameters
$a_i$. Note that the product 
\begin{equation}
   e^{-u/b(m_b)}\,e^{5u/3}
   = \left( \frac{e^{5/6}\LMS}{m_b} \right)^{2u}
   \equiv \left( \frac{\Lambda_V}{m_b} \right)^{2u}
\end{equation}
is RG invariant. The constant $5/6$ arises in the calculation of the 
fermion-loop contribution to the gluon self-energy in the \MS scheme and
compensates the scheme dependence of $\LMS$. We will thus express our 
results in terms of the invariant scale parameter $\Lambda_V$.

If the light-cone distribution amplitudes are given by the model used 
in the explicit calculations of Section~\ref{subsec:asym}, then all of 
the Borel transforms have a leading pole at $u=1$. For the 
hard-spectator term this leads to a power correction of order 
$\Lambda_V/m_b$, which is expected, because the effective scale in the 
hard-spectator terms is not $m_b^2$ but $\mu_h^2\simeq\LQCD m_b$ 
\cite{BBNS3}. In all other cases the leading power correction is of 
order $(\Lambda_V/m_b)^2$. Expanding the Borel transforms about the 
positions of the poles, and including terms up to second order in
the heavy-quark expansion, we find
\begin{eqnarray}\label{powercorrections}
   \Delta a_2 &=& \frac{12C_F}{N\bo}\,\frac{\Lambda_V^2}{m_b^2}\,
    \left( \ln\frac{m_b}{\Lambda_V} - \frac14 - i\pi \right) \nonumber\\
   &&\mbox{}+ \frac{12C_F}{N\bo}\,
    \frac{6\pi^2}{N}\,\frac{f_B f_K}{m_B\lambda_B\,F_0^{B\to K}(0)}
    \nonumber\\
   &&\quad\times
    \Bigg\{ \frac{\Lambda_H}{m_b} 
    \left( \ln\frac{m_b}{\Lambda_H} - \gamma_E - 1 \right)
    + \frac{\Lambda_H^2}{m_b^2}
    \left( \ln\frac{m_b}{\Lambda_H} - \gamma_E + 2 \right) \Bigg\}
    + \dots \,, \nonumber\\
   \Delta a_4^p &=& \frac{12C_F}{N\bo}\,\frac{\Lambda_V^2}{m_b^2}\,
    C_{8g}^{(0)} + \dots \,, \nonumber\\
   \Delta a_6^c &=& \frac{2C_f}{N\bo}\,\frac{\Lambda_V^2}{m_b^2}
    \left\{ C_{8g}^{(0)} + \frac13 \left( 2\ln\frac{m_b}{m_c} -1 \right)
    - \left( 2\ln\frac{m_b}{\Lambda_V} + \frac53 \right) C_4^{(1)}(m_b)
    \right\} + \dots \,, \nonumber\\
   \Delta a_6^u &=& \frac{2C_f}{N\bo}\,\frac{\Lambda_V^2}{m_b^2}
    \left\{ C_{8g}^{(0)}
    + \left( 2\ln\frac{m_b}{\Lambda_V} + \frac53 \right)
    \left[ \frac13 - C_4^{(1)}(m_b) \right] - \frac13 \right\}
    + \dots \,,
\end{eqnarray}
where $\Lambda_H=(m_B/m_b)(\Lambda_V^2/\lambda_B)=O(\LQCD)$ in the 
hard-spectator contribution to $\Delta a_2$. These expressions may serve 
as a model for the leading power corrections to the decay amplitudes 
arising from ``non-factorizable'' soft gluon exchange. Because most of
the corrections are suppressed by two powers of $\Lambda_V/m_b$, their 
numerical effects turn out to be rather small. Only $a_2$ receives a 
first-order correction that can be sizeable. However, the effect of that 
correction can be absorbed into a redefinition of the hadronic parameter 
$\lambda_B$, which in any case is poorly determined theoretically. 
Therefore, power corrections from soft gluon exchange have a minor
impact on the phenomenological analysis of charmless hadronic $B$ 
decays in the framework of QCD factorization.

Let us briefly discuss the changes that occur if the light-cone 
distribution amplitudes are allowed to have an arbitrary endpoint 
behavior, such as 
\begin{equation}
   \Phi_M(x)\simeq {\bar x}^{\delta_M} \Big[ 1 + O({\bar x}) \Big]
\end{equation} 
as $x\to 1$ (and similarly for $x\to 0$). The model used above 
corresponds to $\delta_\pi,\delta_K,\delta_B=1$ and $\delta_K^p=0$. 
These results would continue to hold at any fixed order in a Gegenbauer
expansion of the distribution amplitudes, but may be altered if an 
infinite number of Gegenbauer moments is included. The effect of a more 
general endpoint behavior on the position of the poles in the Borel 
plane depends on the function being considered. As seen from (\ref{sV}), 
the Borel transform corresponding to the vertex function $V_M$ has a 
pole at $u=1$ regardless of the endpoint behavior of the distribution 
amplitude of the meson $M$. For the penguin and hard-spectator terms, 
on the other hand, changing the distribution amplitudes will change the 
position of the leading IR renormalon poles. For the penguin terms the 
modifications are simple: the leading poles in the twist-2 and twist-3 
penguin contributions are shifted from $u=1$ to $u=\delta_K$ (twist-2) 
and $u=1+\delta_K^p$ (twist-3), respectively. The effects on the 
hard-scattering term are slightly more involved. If the $B$-meson and 
kaon distribution amplitudes have the same endpoint behavior, then 
there are both single and double poles at $u=\delta_K=\delta_B$. On the 
other hand, if $\delta_B$ differs from $\delta_K$, then the position of 
the leading simple pole is determined by whichever is smaller.

\boldmath
\subsection{Fermion loops beyond the large-$\bo$ approximation}
\unboldmath
\label{sec:mod}

Whereas in most applications of the large-$\bo$ limit the replacement 
$n_f\to-3\bo/2$ after the calculation of fermion bubble graphs can be
motivated by a diagrammatic analysis, this replacement is hard to 
justify for the penguin coefficients entering the analysis of 
non-leptonic $B$ decays. The fermion-loop contributions proportional to 
$n_f$ are usually accompanied by gauge boson and ghost loops. 
Calculations using bubble resummation are often good approximations to 
full QCD because the replacement $n_f\to-3\bo/2$ (approximately) 
accounts for the extra diagrams not related to fermion bubbles. The 
factor of $n_f$ associated with the penguin coefficients is not of this 
nature. It arises from the contraction of two fermion fields of a 
4-quark operator in the effective weak Hamiltonian. Such fermion loops 
are not accompanied in an obvious way by other QCD loop
diagrams.\footnote{Such diagrams would appear at two-loop order when one
calculates matrix elements of the chromo-magnetic dipole operator. We
have performed a partial analysis of such higher-order contributions and
found their effects to be  strongly suppressed.}
To produce reasonable numerical results for the Borel-resummed kernels 
will require moving beyond the large-$\bo$ approximation. 

A proper counting of fermionic penguin contractions can be achieved 
{\em a posteriori\/} by making the replacement
\begin{equation}\label{repl}
   \frac32\,C_4^{(1)}(m_b)\,G(0,\bar y)
   \to - \frac{C_4^{(1)}(m_b)}{\beta_0}\,
   \Big[ (n_f-2)\,G(0,\bar y) + G(s_c,\bar y) + G(1,\bar y) \Big]
\end{equation}
in the expressions for the Borel transforms $S_{P,2}^p$ and $S_{P,3}^p$ 
in (\ref{SK23}). Upon setting $n_f=5$, (\ref{repl}) accounts for loop 
contributions from three massless quarks, the charm quark, and the 
bottom quark (cf.\ eqs.~(49) and (54) in \cite{BBNS3}). Formally, the 
two sides of (\ref{repl}) coincide in the large-$\bo$ limit, and so 
making this replacement is equivalent to adding some higher-order terms 
to the renormalon calculus. Since there are other terms of subleading 
order in the large-$\bo$ limit not taken into account, this procedure 
may seem arbitrary. Nevertheless, the above replacement is physically 
motivated and, in particular, reproduces the proper imaginary parts and 
heavy-quark mass dependence of the fermion loops. From now on, we will 
refer to the scheme defined by (\ref{repl}) as the ``modified'' 
large-$\bo$ limit.

A replacement analogous to (\ref{repl}) must be made in (\ref{SK23asy}).
The resulting expressions for the leading IR renormalon ambiguities in
the coefficients $a_6^p$ also change. In the last two equations in 
(\ref{powercorrections}), we must replace
\begin{equation}
    - \left( 2\ln\frac{m_b}{\Lambda_V} + \frac53 \right) C_4^{(1)}(m_b)
    \to \frac{2C_4^{(1)}(m_b)}{3\beta_0} \left[ 
    (n_f-2) \left( 2\ln\frac{m_b}{\Lambda_V} + \frac53 \right)
    + 2\ln\frac{m_b}{m_c} \right] .
\end{equation}

\section{Numerical results}
\label{numerical}

In addition to  providing a model for power corrections, the results of 
the previous sections allow us to study the numerical significance of 
higher-order perturbative contributions to the hard-scattering kernels
in the large-$\bo$ limit, which may be considered as a toy model for the
asymptotic behavior of perturbation theory. Partial perturbation series 
for the quantities $a_i$, including terms up to order 
$\beta_0^{n-1}\alpha_s^n$, are obtained by retaining only terms up
to $O(u^{n-1})$ in the Taylor expansions for the Borel transforms about 
$u=0$. The resummed series (in the large-$\bo$ limit) are obtained by
performing the Borel integrals exactly, regularizing the singularities
along the integration contour using the principal value prescription.
The sum of the residues of the poles at half-integer values of $u$ 
provides an estimate of the irreducible ambiguity in the definition of
the resummed perturbation series. Whereas only the leading poles were
considered in the previous section, exact expressions for the sum of
all residues are used in our numerical analysis. 

In Table~\ref{tab-numai}, we show different perturbative approximations
to the coefficients $a_2$ and $(a_4^p+r_\chi\,a_6^p)$ entering the
$B\to\pi\pi,\,\pi K$ decay amplitudes. The first row gives the results 
obtained using the standard QCD factorization approach described in 
\cite{BBNS3}. The following rows show the results obtained in the 
large-$\bo$ limit, at one-loop and two-loop order, and from a numerical 
evaluation of the Borel integrals. The last row contains the renormalon 
ambiguity. For the penguin coefficients, the default entries refer to 
the modified large-$\bo$ limit, whereas results obtained in the strict 
large-$\bo$ limit are given in square brackets. The only sizeable 
renormalon ambiguity arises from the hard-spectator term, which 
contributes to the real part of $a_2$ and corresponds to a first-order 
power correction from soft gluon exchange with the spectator quark. The 
ambiguity in the penguin coefficients is of order $(\LQCD/m_b)^2$ and 
numerically very small. Based on our renormalon analysis, we thus find 
no evidence for a large dynamical enhancement of penguin amplitudes, in 
contrast with some recent conjectures 
\cite{Keum:2001ph,Ciuchini:1997hb}.

\begin{table}
\centerline{\parbox{15cm}{\caption{\label{tab-numai}
Numerical results for the combinations of $a_i$ parameters entering the 
$B\to\pi\pi$ decay amplitudes. Input parameters are $f_\pi=131$\,MeV, 
$f_B=180$\,MeV, $F_0^{B\to\pi}(0)=0.28$, $\lambda_B=350$\,MeV, 
$m_b=4.2$\,GeV, $m_c=1.3$\,GeV, and $r_\chi=1.197$. Perturbative results
refer to the scale $\mu=m_b$ in the \MS scheme except for the 
hard-spectator contribution to $a_2$, where the scale is 
$\mu=\sqrt{\Lambda_h m_b}$ with $\Lambda_h=0.5$\,GeV 
\protect\cite{BBNS3}. Borel resummed results are scale independent.  
Values given in brackets refer to the strict large-$\bo$ limit.}}}
\vspace{0.2cm}
\begin{center}
\begin{tabular}{||c|c|c|c|c|c||}
\hline\hline
 & $a_2$ & $a_4^u+r_\chi\,a_6^u$ & $a_4^c+r_\chi\,a_6^c$ \\ 
\hline
NLO & $0.086-0.084i$ & $-0.081-0.034i$ & $-0.091-0.014i$ \\ 
\hline
1-loop & $0.084-0.075i$ & $-0.082-0.026i$ & $-0.091-0.007i$ \\[-0.15cm] 
 & & $[-0.121-0.071i]$ & $[-0.130-0.053i]$ \\ 
\hline
2-loop & $0.099-0.112i$ & $-0.072-0.036i$ & $-0.090-0.013i$ \\[-0.15cm] 
 & & $[-0.107-0.114i]$ & $[-0.125-0.091i]$ \\ 
\hline
Borel sum & $0.094-0.168i$ & $-0.062-0.031i$ & $-0.088-0.012i$
 \\[-0.15cm]  
 & & $[-0.070-0.115i]$ & $[-0.096-0.096i]$ \\ 
Ambiguity & $0.058-0.008i$ & $0.0004~$ $[0.002]$
 & $-0.0004~$ $[0.001]$ \\  
\hline\hline
\end{tabular}
\end{center}
\end{table}

In regards to  the numerical significance of higher-order perturbative 
corrections, the most striking effect is the large increase of the
imaginary part of the coefficient $a_2$ as one goes to higher  orders in
perturbation theory. An analysis using momentum distribution functions 
shows that the imaginary part of the hard-scattering kernel contributing 
to $a_2$ is dominated by contributions from scales of order a GeV, 
indicating that it can receive large higher-order corrections \cite{BNP}.
The corresponding strong-interaction phase may therefore be larger than
that obtained from the NLO analysis.

In the strict large-$\bo$ limit (shown by the numbers in brackets), the 
penguin coefficients $a_4^p$ and $a_6^p$ mirror the behavior of $a_2$. 
On the other hand, when the physical value $n_f=5$ is used as in 
(\ref{repl}), the phases are reasonably close to those at NLO and do 
not receive large higher-order corrections. This shows that after the 
replacement in (\ref{repl}) there is a strong cancellation between the 
different terms in (\ref{SK23}), which is maintained at higher orders. 
While the  replacement in (\ref{repl}) was motivated on physical 
grounds, it is an open and important question whether such a delicate 
cancellation will persist once the full set of NNLO corrections is 
included in the theory of QCD factorization.

\section{Phenomenological applications}
\label{phenom}

This section addresses the effects of renormalon ambiguities and 
higher-order corrections on the phenomenology of $B$ decays into two 
light pseudoscalar mesons. The  results of Section~\ref{numerical} have 
shown that the most notable effects are the ambiguity in the 
hard-spectator term and an increase in the imaginary parts of the 
kernels at higher orders in perturbation theory. We limit our 
phenomenological analysis to quantities that highlight these effects in 
a transparent way.   

The $B^\pm\to\pi^\pm\pi^0$ decay amplitudes depend strongly on $a_2$ 
(see (\ref{Bpipi})) and are therefore a good probe of the renormalon 
ambiguity in the hard-spectator term. Using the data in 
Table~\ref{tab-numai}, and taking $|V_{ub}/V_{cb}|=0.085$ and 
$|V_{cb}|=0.041$, we find for the CP-averaged branching ratio
\begin{equation}\label{br1}
   10^6\,\mbox{Br}(B^\pm\to\pi^\pm\pi^0)
   = (4.57;\,\,4.73;\,\,4.74\pm 0.50) \,,
\end{equation}
where the numbers in parentheses refer to one-loop order, two-loop 
order, and Borel-resummation, respectively. The NLO result obtained 
using QCD factorization is 4.80. (The central value quoted in 
\cite{BBNS3} is about 10\% larger due to the inclusion of electroweak
penguin contributions.) The magnitude of the renormalon ambiguity may
be compared with the error estimate given in \cite{BBNS3}, where an
uncertainty of ${}^{+0.8}_{-0.4}\,(\mbox{pars.})\pm 0.3\,(\mbox{power})$
was assigned to account for the effects of parameter variations and 
power corrections from higher-twist distribution amplitudes. Since this 
is the mode that is the most vulnerable to the ambiguity in $a_2$, we 
are led to the conclusion that no large power corrections from soft
gluon exchange are present in $B\to\pi\pi,\,\pi K$ decays.

\begin{figure}[t]
\centerline{\epsfxsize=16cm\epsffile{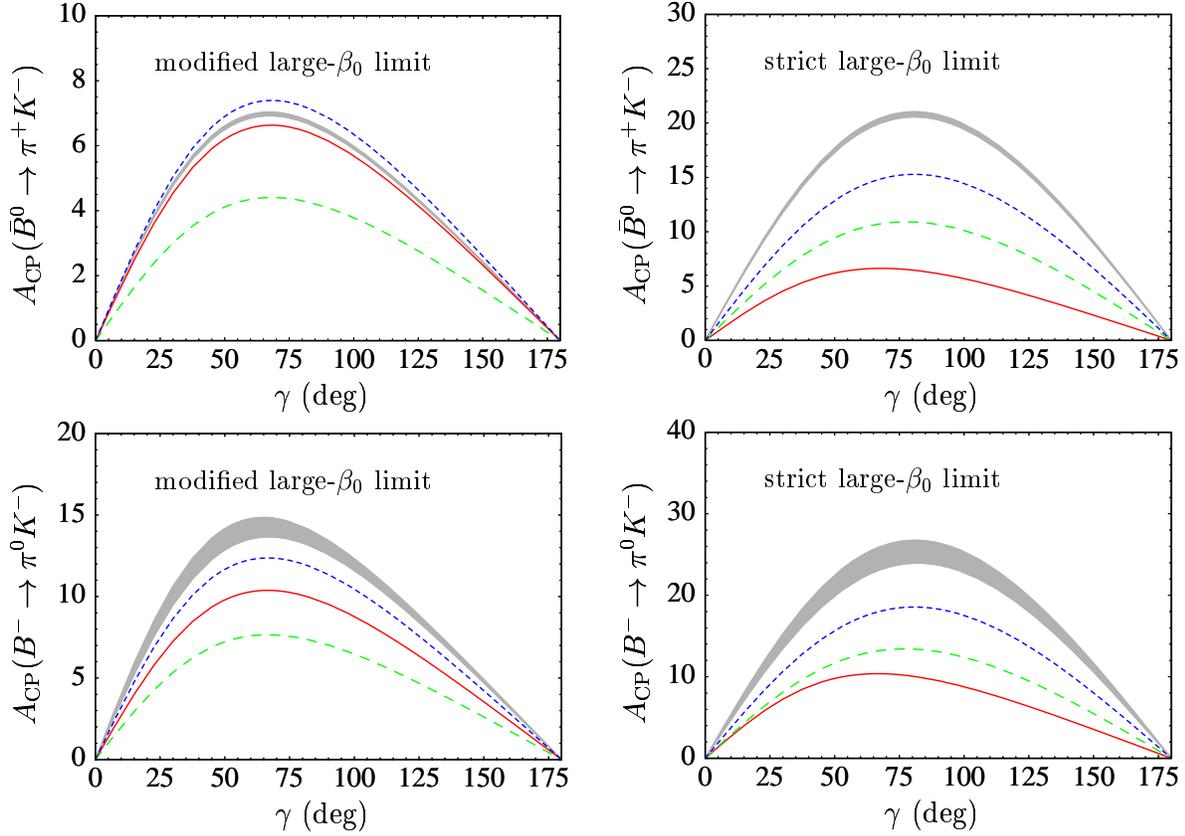}}
\centerline{\parbox{15cm}{\caption{\label{fig:Plots}
Predictions for the $B\to\pi^\pm K^\mp$ and $B^\mp\to\pi^0 K^\mp$ CP
asymmetries in the strict (right) and the modified (left) large-$\bo$ 
limit. The different curves refer to the results at NLO (solid), 
one-loop order (long dashed), two-loop order (short dashed), and 
Borel-resummed (gray bands). For the Borel-resummed results, the width 
of the bands shows the renormalon ambiguity.}}}
\end{figure}

The CP asymmetries of certain decays can be very sensitive to the
imaginary parts of the hard-scattering kernels and  show  clearly the
effects of Borel resummation. As examples, we discuss the asymmetries 
for $B\to\pi^\pm K^\mp$ and $B\to\pi^0 K^\mp$, which to first 
approximation are given by
\begin{eqnarray}
   A_{\rm CP}(\bar B^0\to\pi^+ K^-)
   &\propto& -\sin\gamma\cdot\mbox{Im}(a_4^c+r_\chi\,a_6^c) \,,
    \nonumber\\
   A_{\rm CP}(B^-\to\pi^0 K^-)
   &\propto& -\sin\gamma\cdot\mbox{Im}\left(
    \frac{a_4^c+r_\chi\,a_6^c}{1+R_{\pi K}\,a_2} \right) ,
\end{eqnarray}
where $R_{\pi K}\simeq F_0^{B\to K}(0)/F_0^{B\to\pi}(0)\times f_\pi/f_K
\approx 0.9$. Using asymptotic distribution amplitudes, the $a_i$ 
parameters for $B\to\pi K$ decays are identical to those for 
$B\to\pi\pi$. In Figure~\ref{fig:Plots}, results for the two CP 
asymmetries are shown both in the strict large-$\beta_0$ limit (right) 
and in the modified scheme discussed in Section~\ref{sec:mod}. In both 
schemes, the asymmetries increase strongly at higher orders in 
perturbation theory, such that the Borel-resummed results are almost a 
factor of two larger than the results found at one-loop order. The 
two-loop contributions appear to be the most important higher-order 
effects. This trend suggests that the asymmetries predicted from exact 
NLO calculations in \cite{BBNS3} will become larger upon the inclusion 
of higher-order terms. (As concerns the magnitude of the CP 
asymmetries, we believe that the true values will be somewhere in 
between the two schemes.) It is also apparent from the figure that, in 
all cases, higher-order perturbative corrections are much more important 
than the power corrections suggested by the values of the renormalon 
ambiguities. This provides additional motivation for calculating the 
imaginary part of the hard-scattering kernels at two-loop order. In 
fact, the two-loop imaginary parts (combined with the one-loop real 
parts) of the decay amplitudes are required for a consistent 
description of the CP asymmetries at NLO in RG-improved perturbation 
theory.

\section{Conclusions}

In this paper we have used the renormalon calculus and the large-$\bo$ 
limit of QCD to develop a toy model for higher-order corrections to the 
QCD factorization approach 
as applied to $B$ decays into two light pseudoscalar mesons. We have 
presented novel results for the Wilson coefficients in the large-$\bo$ 
limit, and have used them to show that the RG invariance of the 
hard-scattering kernels entering the QCD factorization formula is 
preserved under Borel resummation. Both the Wilson coefficients and the 
coefficient functions $a_i$ appearing in the amplitudes for 
$B\to\pi\pi,\,\pi K$ decays simplify greatly in the large-$\bo$ limit, 
but still allow for a realistic study of power corrections and 
higher-order perturbative contributions to QCD factorization. We have
presented exact expressions for the Borel transforms of the 
hard-scattering kernels in the factorization formula, from which is it 
possible to derive the terms of order $\bo^{n-1}\alpha_s^n$ in the 
perturbative expansions of the kernels. We have analysed the power 
corrections from soft gluon exchange and obtained a simple model for
the first and second-order corrections to the parameters $a_i$. The 
only numerically significant power correction arises from soft gluon
exchange with the spectator quark, which introduces an error in the 
CP-averaged $B^\pm\to\pi^\pm\pi^0$ branching ratio of about ten percent.
This is roughly the same as the errors associated with the hadronic 
uncertainties and power corrections that were estimated in \cite{BBNS3}. 
The power corrections to the real part of the penguin coefficients are 
of order $(\LQCD/m_b)^2$, and those to the imaginary part are zero 
within the large-$\bo$ limit, which offers numerical evidence against 
models proposing dynamically enhanced penguins.
 
A recurring result of our study is that the imaginary parts of the
hard-scattering kernels may be larger than expected from previous
applications of QCD factorization. A calculation of the imaginary part 
of the kernels at two-loop order would be highly desirable, since it 
would give an answer to the question of whether there are large 
multi-loop contributions to these decays that are not captured in the 
large-$\bo$ limit.

\subsection*{Acknowledgements}

We would like to thank Thomas Becher and Martin Beneke for useful 
discussions. This work was supported in part by the National Science 
Foundation.

\end{document}